\documentclass[onecolumn,a4paper,amsmath,amssymb,floatfix,showpacs]{revtex4}
\usepackage{braket}
\usepackage{graphicx}
\usepackage{dcolumn}
\usepackage{epstopdf}

\newcommand{\Op}[1]{\boldsymbol{\mathsf{\hat{#1}}}}

\newcommand{\gOp}[1]{\boldsymbol{\hat{#1}}}

\def\openone{\leavevmode\hbox{\small1\kern-3.3pt\normalsize1}}

\begin{document}

\title{Cooling molecular vibrations with shaped laser pulses:\\
  Optimal control theory exploiting the timescale separation \\between
  coherent excitation and spontaneous emission}

\author{Daniel M. Reich}
\affiliation{Theoretische Physik,  Universit\"at Kassel,
  Heinrich-Plett-Str. 40, 34132 Kassel, Germany}

\author{Christiane P. Koch}
\affiliation{Theoretische Physik, Universit\"at Kassel,
  Heinrich-Plett-Str. 40, 34132 Kassel, Germany}

\date{\today}

\begin{abstract}
  Laser cooling of molecules employing broadband optical pumping involves a 
  timescale separation between laser excitation and spontaneous
  emission. Here, we optimize the optical pumping step using shaped laser
  pulses. We derive two optimization functionals to
  drive population into those excited state levels that have
  the largest spontaneous emission rates to the target state. We show
  that, when using optimal control, laser cooling of molecules works even if
  the Franck-Condon map governing the transitions is preferential to
  heating rather than cooling. Our optimization functional is also 
  applicable to the laser cooling of other degrees of freedom provided
  the cooling cycle consists of coherent excitation and dissipative
  deexcitation steps whose timescales are separated.  
\end{abstract}

\maketitle

\section{Introduction}
\label{sec:intro}

Laser cooling of atoms or molecules relies on the repeated excitation
and spontaneous emission of light~\cite{CohenAdvances}. When 
the atom or molecule reaches a dark state, i.e., a state
that does not interact with the laser light, it escapes from
the cooling cycle. If 
this occurs before the particle is sufficiently cooled, repumping is
required.  The presence of too many levels that act as dark
states has prevented laser cooling to work for most molecular 
species. However, dark states can also be used to an advantage in
laser cooling when they are populated only by the cooled particles. 
This is utilized for example in subrecoil cooling based on 
velocity selective coherent population trapping~\cite{AspectPRL88}. 
Dark states also play a crucial role in the laser cooling of internal
degrees of freedom~\cite{AllonJCP93,AllonJCP97,AllonCP01}. The
presence of many internal levels requires a broadband optical
excitation which can be realized by femtosecond laser pulses. Cooling
occurs if the target level is populated by spontaneous emission but
remains dark to the laser pulse~\cite{AllonJCP97,AllonCP01}. The dark
state can be realized by destructive interference or simply by 
removing the frequency components corresponding to excitation of the
target level. The latter has recently been realized experimentally,
resulting in successful demonstration of laser cooling of
vibrations~\cite{PilletSci08,ViteauFaraday09,SofikitisNJP09,SofikitisMolPhys10,LignierPCCP11,HorchaniPRA12,WakimOE12}. An extension to cooling rotations is feasible as
well~\cite{OdomPCCP11,ManaiPRL12,ManaiMolPhys13}.

\begin{figure}[b]
  \centering
  \includegraphics[width=0.55\linewidth]{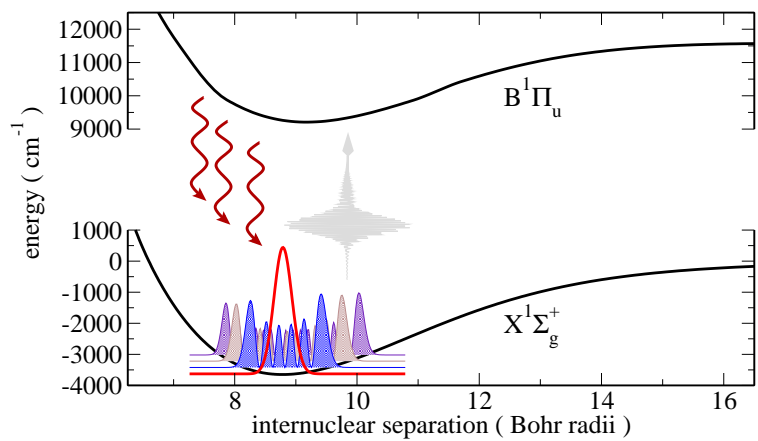}
  \caption{%
    Potential energy curves of the Cs$_2$ electronic states employed
    for the vibrational cooling by optimized optical pumping and
    spontaneous emission. The vibrational ground state (red
    solid curve) is the target state of the optimization,
    vibrationally excited states (shown here $v=5,10,15$) make up the
    initial incoherent ensemble.  
  }
  \label{fig:pots_cooling}
\end{figure}
In the experiments of
Refs.~\cite{PilletSci08,ViteauFaraday09,SofikitisNJP09,SofikitisMolPhys10,LignierPCCP11,HorchaniPRA12,ManaiPRL12,ManaiMolPhys13},
cooling the internal degrees of freedom by broadband optical pumping
was preceded by standard laser cooling of atoms to temperatures of the
order of 100$\,\mu$K and then photoassociating the atoms into weakly
bound excited state molecules. Photoassociation~\cite{FrancoiseReview,JonesRMP06} is
followed by spontaneous emission, yielding molecules in the ground
electronic  state. Depending on the choice of excited state potential, a
significant part of the molecules might end up in ground 
state levels with comparatively small vibrational quantum
numbers~\cite{PilletSci08,LignierPCCP11}.  These molecules can be
laser cooled by broadband optical pumping as
illustrated in Fig.~\ref{fig:pots_cooling}: An incoherent ensemble
of molecules in different vibrational levels of the electronic ground
state is excited by a broadband laser pulse to an electronically
excited state.
The electronically excited molecules will decay by spontaneous
emission back to the ground state. The branching ratio for the
different ground state vibrational levels is determined by
the Franck-Condon factors or, more precisely, transition matrix
elements, between ground and excited state levels. Some decay will 
always lead to the ground vibrational level.
Repeated broadband optical pumping then accumulates 
the molecules in the ground vibrational level~\cite{PilletSci08}. 

The overall cooling rate is determined by the timescale of the
dissipative step, i.e., the spontaneous
emission lifetime~\cite{AllonJCP93,AllonJCP97,AllonCP01}. It cannot be
modified by the coherent interaction of the molecules with the laser
pulse. However, the pulses can be shaped such as to populate those
excited state levels which preferentially decay into the
target level. Here we show that this minimizes the number of required
optical pumping cycles. 
Moreover, we demonstrate that optimal pulse shapes allow for cooling
even in cases where 
the Franck-Condon map is preferential to heating rather than
cooling. This is the case when the excited state levels show similar
Einstein coefficients for many ground state vibrational levels. Rather
than accumulating the molecules in a single target level, spontaneous
emission then distributes the population incoherently over many
levels, effectively heating the molecules up.

We employ optimal control theory to calculate the pulse
shapes. Instead of treating the full dissipative dynamics of the
excitation/spontaneous emission cycle, we take advantage of the
timescale separation between the coherent interaction of the molecules
with the laser pulse, on the order of 10$\,$ps, and the spontaneous
decay with excited state lifetimes of the order of 10$\,$ns. Seeking a
pulse that populates those excited state levels with the largest
Einstein coefficients with the target ground state level allows us to
treat the decay implicitly. We formulate two optimization functionals
that are independent of the specific initial state. Thus we
obtain an optimized pulse shape that remains unchanged over the
complete cooling process consisting of many repeated
excitation/spontaneous emission cycles. The two optimization
functionals realize different cooling mechanisms: One is based on
optical pumping from all thermally populated ground state levels
symmetrically, whereas the other one forces the thermally populated
ground state levels into an 'assembly line'. Only the first level in
the line is transferred to the excited state while population from all
other levels is reshuffled, one after the other into the first level, via Raman
transitions. This suppresses heating actively and
allows us to answer the question of what is the fundamental requirement
of the molecular structure to allow for cooling. 

Our paper is organized as follows. Section~\ref{sec:model} introduces
our model for the interaction of the molecules with the laser pulse
and the spontaneous emission. We derive the optimization functionals
for cooling in Sec.~\ref{sec:functional} and present our numerical
results in Sec.~\ref{sec:results}, comparing vibrational laser cooling for
Cs$_2$ and LiCs molecules. We conclude in Sec.~\ref{sec:concl}.

\section{Model}
\label{sec:model}

We consider Cs$_2$ and LiCs molecules in their electronic ground
state after photoassociation and subsequent spontaneous emission. 
The excited state for optical pumping is chosen to be the
$B^1\Pi_u$ state as in the experiment for Cs$_2$ molecules of
Refs.~\cite{PilletSci08,ViteauFaraday09,SofikitisNJP09,SofikitisMolPhys10}.  
This state is comparatively isolated such that population leakage to other
electronic states due to e.g. spin-orbit interaction is minimal. 
The Hamiltonian describing the interaction of the molecules with
shaped femtosecond laser pulses in the rotating-wave approximation reads
\begin{equation}
  \label{eq:H_Cs2}
  \Op H =
  \begin{pmatrix}
    \Op T + V_{X^1\Sigma^+}(\Op R) & \frac{1}{2}\epsilon^*(t)\, \Op\mu \\
    \frac{1}{2}\epsilon(t)\, \Op\mu  &     \Op T + V_{B^1\Pi}(\Op R) - \omega_L
  \end{pmatrix}\,,
\end{equation}
where $\Op T$ denotes the vibrational kinetic energy.
$V_g=V_{X^1\Sigma^+}(\Op R)$ and  $V_e=V_{B^1\Pi}(\Op R)$ are the potential
energy curves as a function of interatomic separation, $\Op R$,
of the electronic ground and excited state (note that for
Cs$_2$ the $X$ state is of gerade symmetry and the $B$ state of
ungerade symmetry). $\Op\mu$ is the transition dipole moment,
approximated here to be independent of $\Op R$. The laser pulse is
characterized by its carrier frequency, $\omega_L$, and complex shape,
$\epsilon(t)=|\epsilon(t)|e^{i\phi(t)}$, 
with the time-dependent phase $\phi(t)$ referenced to the phase of
the carrier frequency. The potential energy curves are found in
Refs.~\cite{AmiotJCP02} and \cite{StaanumPRA07} for the electronic
ground state and in Refs.~\cite{DiemerCPL89} 
and \cite{GrocholaJCP09} for the electronically excited
state of Cs$_2$ and LiCs, respectively. 

The decay of the excited state molecules back to the electronic ground
state is described by the spontaneous emission rates,
\begin{equation}
\gamma_{v'J'}^d=\sum_{v''J''} A_{v'J',v''J''}\,.
\end{equation}
The Einstein coefficients $A_{v'J',v''J''}$
are determined by the Franck-Condon factors, 
\begin{equation}
  A_{v'J',v''J''} = \frac{4\alpha^3}{3e^4\hbar^2}
  H_{J'} (E_{v'J'}-E_{v''J''})^3 
  \Big| \langle\varphi^{B}_{v'J'}|
  \Op\mu|\varphi^{X}_{v''J''}\rangle\Big|^2\,,
  \label{eq:Einstein}
\end{equation}
where $H_{J'}$ is the H\"onl-London factor equal to  
$(J'+1)/(2J'+1)$  for $J'=J''-1$ and equal to $J'/(2J'+1)$ for
$J'=J''+1$, $\alpha$ denotes the fine structure constant and
$e$ the electron charge. $|\varphi^{X}_{v''J''}\rangle$ and
$|\varphi^{B}_{v'J'}\rangle$ 
are the rovibrational eigenstates of the $X^1\Sigma^+$ electronic
ground state and the $B^1\Pi$ excited state, respectively. 
We will neglect rotations in the following since the Einstein
coefficients are essentially determined by the Franck-Condon factors,  
$\langle\varphi^{B}_{v'J'}| \Op\mu|\varphi^{X}_{v''J''}\rangle \approx
\langle\varphi^{B}_{v'0}| \Op\mu|\varphi^{X}_{v''0}\rangle$. 

\begin{figure}[tbp]
  \centering
  \includegraphics[width=0.45\linewidth]{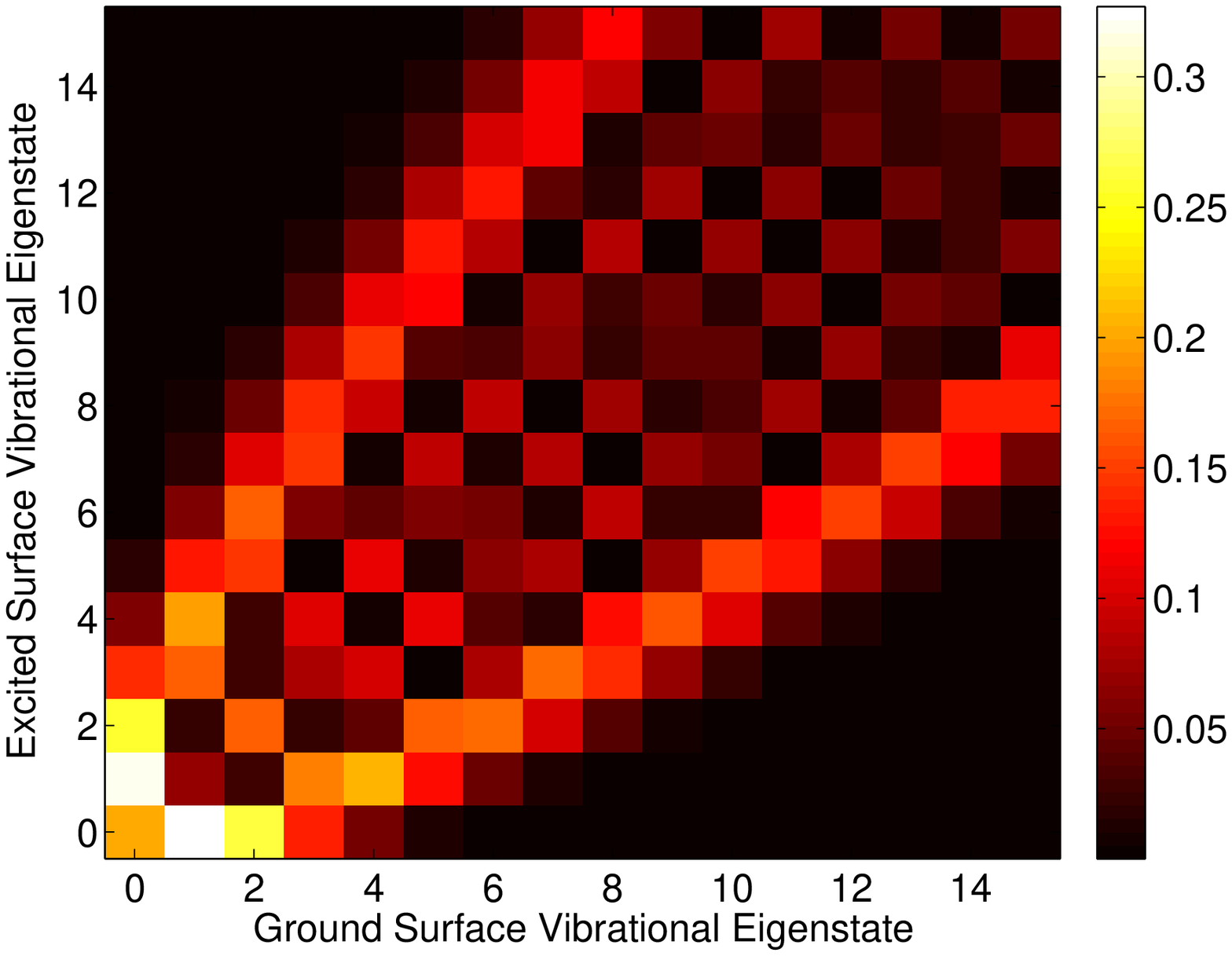}
  \includegraphics[width=0.45\linewidth]{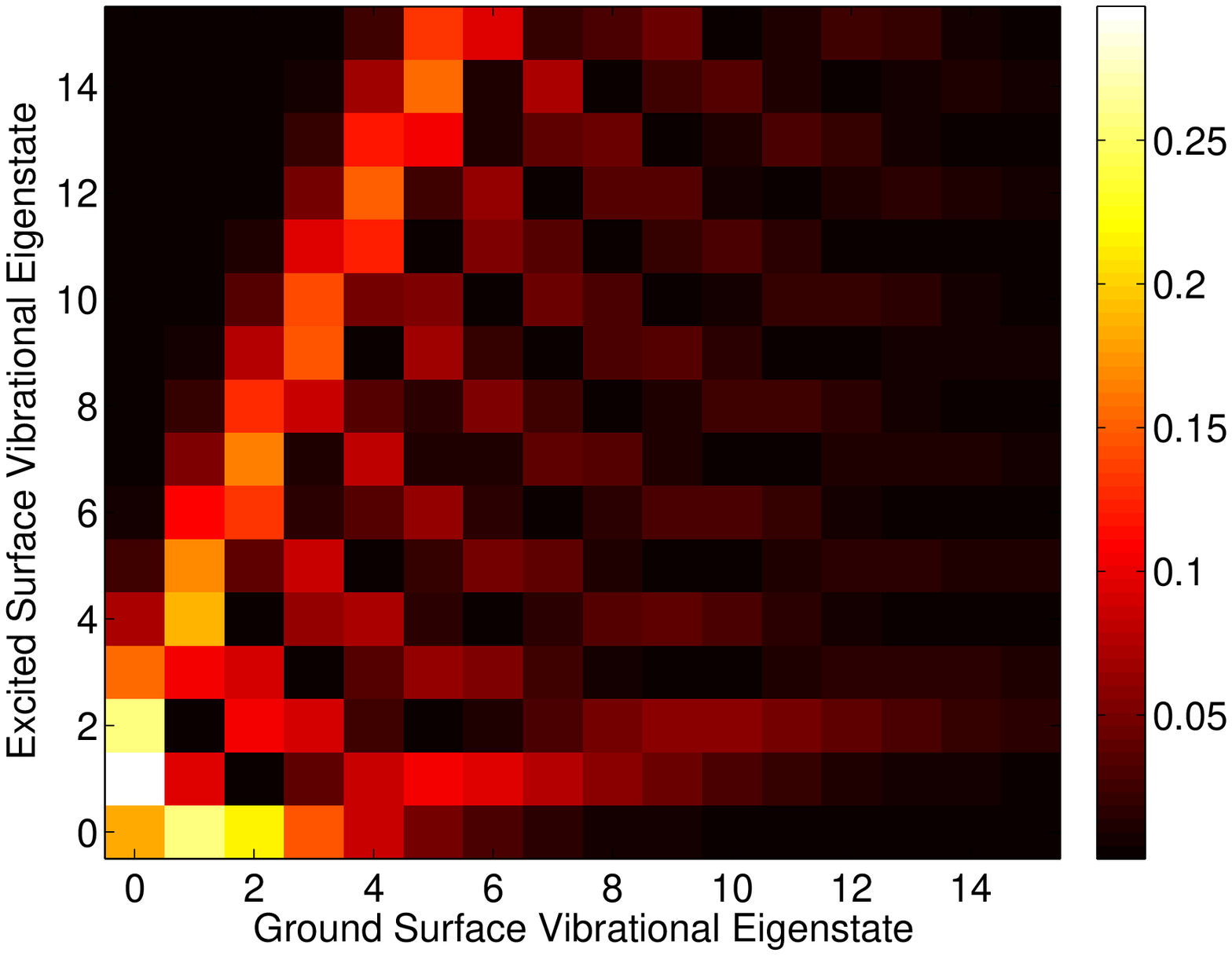}
  \caption{Franck-Condon map, 
    $\langle\varphi^{B}_{v'}| \Op\mu|\varphi^{X}_{v''}\rangle$, 
    as a function of ground and excited state levels, $v''$ and
    $v'$, respectively, for Cs$_2$ (left) and LiCs (right).
    Optical pumping at the right edge of the compact parabola for Cs$_2$
    ensures cooling. This is in contrast to LiCs where absence of a
    compact boundary of the large transition matrix elements implies
    spontaneous emission towards levels with larger $v''$, i.e., 
    heating. 
  }
  \label{fig:FCF}
\end{figure}
Figure~\ref{fig:FCF} displays the Franck-Condon map that governs the
spontaneous emission for Cs$_2$ and LiCs. A compact parabola of large
transition matrix elements is observed for Cs$_2$, cf. left-hand side
of Fig.~\ref{fig:FCF}. Excitation at the
right edge of this parabola can simply be ensured by removing part of
the broadband spectrum~\cite{PilletSci08}. Spontaneous emission then
will occur to levels with $v''\le v_{initial}$, and repeated cycles of
broadband excitation and spontaneous emission results in vibrational
cooling~\cite{PilletSci08}. The situation changes completely for
LiCs, cf. right-hand side of Fig.~\ref{fig:FCF}. There is no compact
boundary separating large from small transition matrix elements, and a
given excited state level has many non-zero transition matrix elements
of similar magnitude. Spontaneous emission will thus spread the
population, and even worse, will do so preferentially
toward levels with $v''\ge v_{initial}$, leading to heating rather
than cooling.

\section{Optimization functional for vibrational cooling 
  of molecules}
\label{sec:functional}

We will employ Krotov's
method~\cite{Konnov99,SklarzPRA02,PalaoPRA03,ReichKochJCP12} to
optimize vibrational cooling of molecules. The total optimization
functional is then split into a final-time target $J_T$ and an
intermediate-time cost $J_t$,
\begin{equation}
  \label{eq:J}
  J = J_T + \int_0^T J_t \,dt\,,
\end{equation}
and will be minimized. 
We choose the intermediate-time cost to minimize the change in pulse
fluence~\cite{PalaoPRA03}, 
\begin{equation}
  \label{eq:J_t}
  J_t = \frac{\lambda}{S(t)}\left[
    \epsilon(t)-\epsilon_{ref}(t)\right]^2\,, 
\end{equation}
where $\lambda$ is a free parameter, $S(t)$ a shape function enforcing the
pulses to be switched on and off smoothly and $\epsilon_{ref}(t)$ a
reference field, taken to be the pulse from the previous iteration. The
final time $T$ is also a free parameter.

We construct $J_T$ such as to avoid solution of the Liouville von
Neumann equation for the density matrix during optimization. This is
possible due to a separation of the timescales for spontaneous decay,
of the order of 10$\,$ns, and the coherent interaction of
the molecules with laser light, of the order of 10$\,$ps. Moreover, it
allows for determining the laser field that is the best possible
compromise, no matter what is the initial state. In other words, the
same pulse can be used over and over again, accumulating molecules in
the target state. We discuss two possible choices for
the final-time functional.

\subsection{Functional for exciting all vibrationally excited ground
  state levels symmetrically}
\label{subsec:J_symm}

The main
idea of this functional is to excite all vibrationally excited ground
state levels symmetrically into those excited state levels which
preferentially decay toward the target state
$|\varphi^g_0\rangle$ while minimizing potential heating. Symmetric
excitation ensures that all ground state levels in the thermal 
ensemble are treated homogeneously. 
The initial state for each laser pulse 
is  given by an unknown incoherent distribution over
ground state vibrational levels,
$|\psi_i(0)\rangle=|\varphi^g_i\rangle$, $i=1,\ldots,n_{max}$. Each of these 
levels is excited by the pulse and subject to the ensuing
dynamics, giving rise to wavepackets $|\psi_i(t)\rangle$ which decay
by spontaneous emission to ground state vibrational levels. 
The spontaneous decay of the excited 
state component of the $i$th wavepacket $|\psi_i(t)\rangle$ to the
target level $|\varphi_0^g\rangle$ is determined by 
the temporally averaged overlap,
\begin{equation}
  \label{eq:J_i}
  \sigma_i = \frac{1}{T_{e}} 
  \int_T^{T+T_{e}}
  \left|\Braket{\psi_{i}(t)|\Op{P}_e\,\gOp\mu|\varphi_0^g}\right|^2 dt \,,
\end{equation}
where $T_e$ denotes the excited state lifetime and $\Op{P}_e$ is the
projector onto the excited electronic state. Shifting the time axis by $-T$,
inserting the completeness relation for vibrational levels on the
excited state and  denoting the Franck-Condon factors 
$\langle\varphi^e_n|\gOp\mu|\varphi^g_m\rangle$  by $\eta_{nm}$,
Eq.~\eqref{eq:J_i} becomes 
\begin{eqnarray*}
  \sigma_i = \frac{1}{T_{e}}\int_0^{T_{e}} \sum_{n,m}
  e^{i(E^e_n-E^e_m)t} \eta_{n0}\eta^*_{m0}
  \langle\psi_i(T)|\varphi^e_n\rangle\langle \varphi^e_m|\psi_i(T)\rangle dt\,,
\end{eqnarray*}
where $E^e_n$ is the eigenenergy corresponding to $|\varphi^e_n\rangle$. 
The integral is readily evaluated, yielding 
\begin{eqnarray*}
  \sigma_i =  \sum_{n \neq m} \frac{1}{iT_e(E^e_n-E^e_m)}
  \left(  e^{i(E^e_n-E^e_m)t} -1\right)
  \eta_{n0}\eta^*_{m0}
  \langle\psi_i(T)|\varphi^e_n\rangle\langle
  \varphi^e_m|\psi_i(T)\rangle
  + \sum_n |  \eta_{n0}|^2 \left|  \langle\psi_i(T)|\varphi^e_n\rangle\right|^2\,.
\end{eqnarray*}
Due to the timescale separation, $1/(T_e (E^e_n-E^e_m))$ is at most of
the order $10^{-4}$, and the temporally averaged overlap is
well approximated by the second term alone, 
\begin{equation}
  \label{eq:sigma_i}
  \sigma_i = \sum_n |  \eta_{n0}|^2 \left| \langle\psi_i(T)|\varphi^e_n\rangle\right|^2\,.
\end{equation}

The timescale separation also allows for neglecting the accidental
creation of coherences in the ground state density matrix after each
cooling cycle.
While the initial ensemble most likely is a completely incoherent
mixture, the state obtained on the ground electronic surface after one
cooling cycle may contain coherences. Accidentally, this could lead to
accumulation of molecules in an undesired dark state, i.e., a certain
coherent superposition of vibrational eigenstates. 
However, the free evolution of the molecule introduces rapidly
oscillating prefactors for each eigenstate. These oscillations are much more
rapid than the time necessary for decay to the ground
surface. Therefore, the system will be in a
superposition of eigenstates with a fixed modulus but random phase
before the next pulse arrives. If necessary, this can be strictly
enforced by introducing a small, randomized waiting period between
cycles. Since a dark state requires a fixed phase
relation, accumulation in the dark state is effectively ruled out. 

Ignoring  coherences, the initial ensemble for each pulse 
is described only in terms of the
vibrational populations, and maximizing the excitation of each
vibrational level corresponds to minimizing 
\begin{equation}
  \label{eq:J_yield}
  J_{yield} = 1 - \sum_{n=1}^{n_{max}} \sigma_n\,. 
\end{equation}
Symmetric excitation of all levels is ensured by balancing the yield
with respect 
to an arbitrarily chosen level out of the initial ensemble, $1\le
n^*\le n_{max}$,
\begin{equation}
  \label{eq:J_sym}
  J_{sym} = \sum_{n=1 (n\neq
    n^*)}^{n_{max}}\left(\sigma_n-\sigma_{n^*}\right)^2\,. 
\end{equation}
$J_{sym}$ is required because otherwise the yield could be maximized
by very efficiently exciting only some levels in the initial
ensemble. This would result in incomplete cooling. 
In addition to efficiently exciting all vibrationally excited ground
state levels, the target state must be kept dark. This is achieved by
enforcing the steady-state condition,
\begin{equation}
  \label{eq:J_ss}
  J_{ss} = 1 -
  \left|\langle\varphi^g_0|\Op U(T,0;\epsilon)|\varphi^g_0\rangle\right|^2\,.
\end{equation}
A further complication arises from the fact that molecules could 
dissociate during the cooling process. This is a source of loss and
needs to be strictly prevented. The most efficient way of enforcing
this requirement is to avoid leakage out of the initial ensemble of
ground state vibrational levels,   
\begin{equation}
  \label{eq:J_leak}
  J_{leak} =  \sum_{m'=n_{max}+1} \sum_{m=0}^{n_{max}}
  \left|\langle\varphi^g_{m'}|\Op
    U(T,0;\epsilon)|\varphi^g_m\rangle\right|^2 +
  \sum_{m'=n_{max}+1}\sum_l\sum_{m=0}^{n_{max}} |\eta_{lm'}|^2 
  \left|\langle\varphi^e_l|\Op
    U(T,0;\epsilon)|\varphi^g_m\rangle\right|^2
  \,.
\end{equation}
The first term in Eq.~\eqref{eq:J_leak} suppresses population
transfer, via Raman transitions, from the initial ground 
state ensemble into higher excited ground state levels, whereas the
second term suppresses population of  excited state levels that have
large Franck-Condon factors with ground state levels outside of the
initial ensemble. 
$J_{leak}$ does not only counter dissociation of the molecules but also
undesired heating. 

The complete final-time functional is given by the multi-objective target
of keeping the target state dark, efficiently exciting all other
vibrational levels in the initial ensemble and avoiding leakage out of
the initial ensemble, 
\begin{equation}
  \label{eq:J_T_symm}
  J_T^{sym} = \lambda_{ss}J_{ss} + \lambda_{leak} J_{leak} + 
  \lambda_{yield} J_{yield} + \lambda_{sym}J_{sym}\,,
\end{equation}
where the $\lambda_j>0$ allow to weight the separate contributions
differently. The functional~\eqref{eq:J_T_symm} will yield optimized
pulses that cool when used in repeated excitation/deexcitation cycles,
unless the molecule under consideration has a Franck-Condon map that
strongly favors heating rather than cooling such that simultaneously
fulfilling all targets imposed by the functional becomes very
difficult. This raises the question of what 
is the minimum requirement on the transition matrix elements to obtain
cooling. It has led us to define a
second optimization functional.

\subsection{Functional for assembly-line cooling}
\label{subsec:J_ass}

The main idea of this functional is to optimize population transfer to
the electronically excited state only for a single ground
state level $n^*$. The excited state levels that are reached from
$n^*$ need to have
Franck-Condon factors that are favorable to cooling (in the extreme
case, a single excited state level with favorable Franck-Condon factor
is sufficient). The population of all other vibrationally excited
ground state levels is simply reshuffled via Raman transitions, 
populating preferentially $n^*$. For example, if the cooling target is
the ground state and  we choose
$n^*=1$, all higher levels are reshuffled into the next lower
level, forming an 'assembly line' which ends in $n=n^*$. 

The corresponding functional contains the steady state and leakage
terms just as Eq.~\eqref{eq:J_T_symm}. The excitation term now targets
only $n^*$, taken to be $n^*=1$,
\begin{equation}
  \label{eq:J_yield_ass}
  \tilde J_{yield} = 1 - \sigma_{1}\,,
\end{equation}
and population reshuffling towards lower vibrational levels is
enforced by the assembly-line term,  
\begin{equation}
  \label{eq:J_ass}
  J_{ass} = 1 - \frac{1}{n_{max}-1}\sum_{n=2}^{n_{max}}
  \left|\langle\varphi^g_{n-1}|\Op
    U(T,0;\epsilon)|\varphi^g_n\rangle\right|^2\,.
\end{equation}
Similarly to Eq.~\eqref{eq:J_T_symm}, the complete final-time
functional for 
assembly line cooling is given by summing all contributions, 
\begin{equation}
  \label{eq:J_T_ass}
  J^{ass}_T = \lambda_{ss}J_{ss} + \lambda_{leak} J_{leak} + 
  \lambda_{yield} \tilde J_{yield} + \lambda_{ass}J_{ass}
\end{equation}
with weights $\lambda_j>0$.
In Eq.~\eqref{eq:J_T_symm}, heating is countered only via the leakage
term, whereas Eq.~\eqref{eq:J_T_ass} avoids it actively.

\subsection{Krotov's method for vibrational cooling}
\label{subsec:Krotov}

The optimization functionals, Eq.~\eqref{eq:J_T_symm} and
Eq.~\eqref{eq:J_T_ass}, represent the starting point for deriving the
coupled control equations that must be solved iteratively to obtain
the optimized pulse. Following Krotov's
method~\cite{ReichKochJCP12}, we obtain a set of three equations with
prescribed discretization for each iteration step $i$:
\begin{itemize}
\item Forward propagation of  each state in the initial thermal
  ensemble according to 
  \begin{equation}
    \label{eq:tdse}
    i\hbar \frac{\partial}{\partial t}|\psi^{(i+1)}_n(t)\rangle = 
    \Op H [\epsilon^{(i+1)}]|\psi^{(i+1)}_n(t)\rangle\,,
    \quad |\psi^{(i+1)}_n(t=0)\rangle = |\varphi^g_n\rangle\,, \quad n=1,\ldots,n_{max}\,,
  \end{equation}
  with $\Op H$ given by Eq.~\eqref{eq:H_Cs2}. 
\item Backward propagation of the adjoint states, 
  \begin{equation}
    \label{eq:btdse}
    i\hbar \frac{\partial}{\partial t}|\chi^{(i)}_n(t)\rangle = 
    \Op H[\epsilon^{(i)}] |\chi^{(i)}_n(t)\rangle \,,
    \quad |\chi^{(i)}_n(t=T)\rangle = 
    \nabla_{\langle\psi_n|}J^{sym/ass}_T\big|_{\{|\psi^{(i)}(T)\rangle\}}
    \quad n=1,\ldots,n_{max}\,,
  \end{equation}
  with the 'initial' condition at time $t=T$ given by the derivatives of the
  final-time functional, 
  Eq.~\eqref{eq:J_T_symm} or Eq.~\eqref{eq:J_T_ass}, with respect to 
  $\langle\psi_n|$, evaluated using the final-time forward propagated
  states,   $|\psi^{(i)}_n(T)\rangle$. 
\item Update of the control by 
  \begin{equation}
    \label{eq:update}
    \epsilon^{(i+1)}(t) = \epsilon^{(i)}(t) + \frac{S(t)}{\lambda}
    \mathfrak{Im}\left\{
    \sum_{n=1}^{n_{max}}\langle\chi_n^{(i)}(t)|\gOp\mu|\psi_n^{(i+1)}(t)\rangle
    + \frac{1}{2}\sigma(t)
    \sum_{n=1}^{n_{max}}\langle\psi_n^{(i+1)}(t)-\psi_n^{(i)}(t)|
    \gOp\mu|\psi_n^{(i+1)}(t)\rangle
    \right\}
  \end{equation}
  with $|\psi_n^{(i+1)}(t)\rangle$, $|\psi_n^{(i)}(t)\rangle$ and
  $|\chi_n^{(i)}(t)\rangle$  solutions of Eqs.~\eqref{eq:tdse}
  and~\eqref{eq:btdse}, respectively.
  $J_T^{sym}$ is a polynomial of fourth
  order in the states, whereas $J_T^{ass}$ is quadratic
  in the states. This means that $J_T^{sym}$ requires the non-linear version
  of Krotov's method, and $\sigma(t)$ is given by
  $\sigma(t)=-(2A+\epsilon_A)$~\cite{ReichKochJCP12}. For
  $J_T^{ass}$, the linear version is sufficient, i.e.,
  $\sigma(t)=0$.   $A$ can be estimated 
  analytically by evaluating a supremum over the second order
  derivatives of $J_T^{sym}$,   and $\epsilon_A$ is a non-negative
  number.  The analytical estimate of $A$ usually is much larger than
  the actual value of $A$ required to ensure monotonicity of the
  algorithm. Since a large value of $A$ slows down convergence, it is
  much better to approximate $A$ numerically, using Eq.~(25) of
  Ref.~\cite{ReichKochJCP12}. 
\end{itemize}
It turned out, however, that the non-convexity of $J^{sym}_T$ is small
in practice, and 
both the linear and the non-linear version of Krotov's method
behave very similarly. This can be rationalized by the fact that only one term in
$J_T^{sym}$,  $J_{sym}$, is non-convex and its impact on the
convergence is small compared to that of the other terms in   $J_T^{sym}$. 
The results presented below were all obtained for $\sigma(t)=0$ in
Eq.~\eqref{eq:update}. 

Instead of the square modulus in the overlaps of
Eqs.~\eqref{eq:sigma_i}, \eqref{eq:J_ss}, \eqref{eq:J_leak} and
\eqref{eq:J_ass}, it is also possible to use the real part of the
overlap~\cite{PalaoPRA03}. 
This sets a global phase which is not neccessary but shows a better
initial convergence for bad guess pulses. 
The latter is due to the specific form of the 'initial' costates,
$|\chi_n^{(i)}(T)\rangle$, 
which remain constant for real part functionals while 
depending linearly on the final-time forward propagated states,
$|\psi_n^{(i)}(T)\rangle$,  for the square modulus functional.
Hence real part functionals cannot take values close to zero leading
to very small gradients as is the case for square modulus
functionals. This is important in particular for the assembly-line
term, for which formulating a 
good guess pulse is difficult,  and our results
presented below were obtained with the real part instead of the square
modulus in Eq.~\eqref{eq:J_ass}.

\section{Optimization results}
\label{sec:results}

We choose our guess pulses such as to avoid small gradients at the
beginning of the optimization. In all examples, they are taken to be Gaussian
transform-limited pulses of moderate intensity with central frequency
and spectral width chosen to excite a number of transitions that are
relevant for the cooling process. The latter are easily read off 
the Franck-Condon matrices in Fig.~\ref{fig:FCF}. 
The choice of the  $\lambda_j$ is determined by the relative
importance of the individual terms in the optimization functionals.
A large value for the steady-state and leakage terms are impedient
since a low value of these functionals will prevent a high
repeatability of the excitation/deexcitation steps, 
effectively reducing the attainable yield. In contrast, 
a slightly lower yield for an individualy step can easily be
amended by few additional cycles. Consequently, as a rule of
thumb, $\lambda_{ss}$ and $\lambda_{leak}$ should be chosen larger
than $\lambda_{yield}$ and $\lambda_{sym}$ or $\lambda_{ass}$, 
respectively. This is more important for the symmetrised cooling since
in the assembly line case the leakage is much easier to prevent
by virtue of the mechanism. Hence it proved in our calculation sufficient
to choose all $\lambda$ equal to one for the assembly line functional
while it proved useful to choose $\lambda_{leak} = \lambda_{sym} = 1$,
$\lambda_{ss} = 2$ and $\lambda_{yield} = 0.4$ for the symmetrised
functional.

\begin{figure}[tbp]
  \centering
  \includegraphics[width=0.5\linewidth]{Fig3}
  \caption{Optimizing the vibrational cooling of Cs$_2$ molecules
    using symmetrized excitation: Value of the total functional,
    Eq.~\eqref{eq:J_T_symm}, and its components vs iterations of the
    optimization algorithm ($n_{max}=10$).} 
  \label{fig:Cs2_symm}
\end{figure}
\begin{figure}[tbp]
  \centering
  \includegraphics[width=0.55\linewidth]{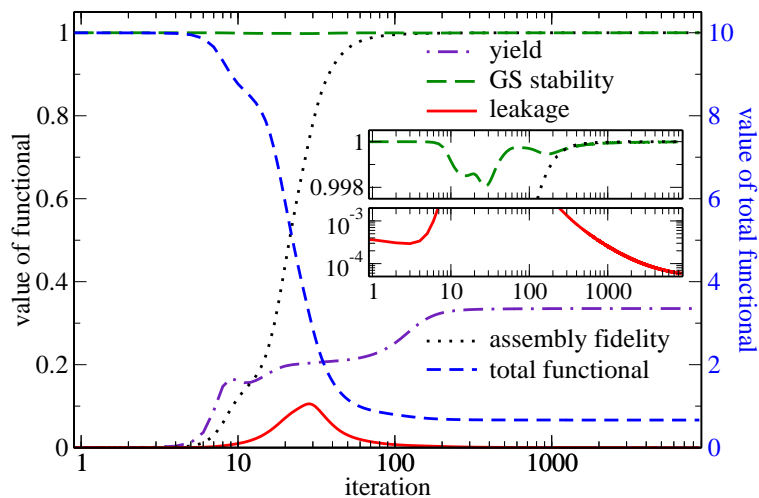}
  \caption{Optimizing the vibrational cooling of Cs$_2$ molecules
    using assembly-line cooling: Value of the total functional,
    Eq.~\eqref{eq:J_T_ass}, and its components vs iterations of the
    optimization algorithm ($n_{max}=10$).} 
  \label{fig:Cs2_ass}
\end{figure}
We first study vibrational cooling of Cs$_2$ molecules, taking
$n_{max}=10$. Due to the
favorable Franck-Condon map, optimization is not required in this case
but helps
to reduce the number of cooling cycles. The behavior of the single
contributions to the optimization functional as well as its
total value are plotted in Fig.~\ref{fig:Cs2_symm}
for $J_T^{sym}$ and in Fig.~\ref{fig:Cs2_ass} for $J_T^{ass}$.
In both cases, monotonous convergence is observed for the total
functional as expected, cf. blue dashed lines in
Figs.~\ref{fig:Cs2_symm} and~\ref{fig:Cs2_ass}. The dark-state
condition for the 
target state is perfectly obeyed for symmetrized excitation
throughout the optimization (green long-dashed line in the inset of 
Fig.~\ref{fig:Cs2_symm}) but presents a slightly more difficult
constraint to fulfill for assembly-line cooling (green long-dashed
line in the inset of Fig.~\ref{fig:Cs2_ass}, note that the stability of
the ground state is given by $1-J_{ss}$).  A final value of 
$1-J_{ss}=9\cdot 10^{-6}$  
ensures also for assembly-line cooling accumulation in the
target state for 10000 cooling 
cycles. This is much more than required as we show below. For
optimization using $J_T^{sym}$, the excitation yield, given by
$1-J_{yield}$,  
measures excitation of all levels in the initial ensemble, and reaches
a value above 0.9, cf. purple dot-dashed line in
Fig.~\ref{fig:Cs2_symm}. 
This together with the fact that the final value of $J_{sym}$ (black
dotted line in Fig.~\ref{fig:Cs2_symm}) is
$10^{-6}$ implies that a pulse that excites all levels in the
initial ensemble with similar efficiency can indeed be found. 
For optimization using $J_T^{ass}$, the excitation yield, $1-\tilde
J_{yield}$, takes a smaller final value (purple dot-dashed line in
Fig.~\ref{fig:Cs2_ass}). This reflects the fact that $1-\tilde
J_{yield}$ measures only excitation out of $v''=1$ and its maximum is
given by 0.335, whereas the
population reshuffling of the other levels is captured by $1-J_{ass}$
(black dotted line in Fig.~\ref{fig:Cs2_ass}). The latter takes a
final value close to one, suggesting that the pulse reshuffles all
higher excited ground state levels in the desired way. 
This indicates efficient excitation at the end of the assembly line as
desired. Thus both optimization functionals, Eq.~\eqref{eq:J_T_symm}
and Eq.~\eqref{eq:J_T_ass}, yield pulses which effectively excite all
higher vibrational levels while keeping the target state dark.  
A striking difference between optimization with $J_T^{sym}$ and
$J_T^{ass}$ is found only in the ability of the optimized pulses to
suppress leakage out of the initial ensemble (red solid lines in 
Fig.~\ref{fig:Cs2_symm} and~\ref{fig:Cs2_ass}). While $J_{leak}$ takes
a final value of about 0.014 for symmetrized excitation, it can be
made smaller than $10^{-4}$ for assembly line cooling. In the latter
case, $J_{leak}$ could be further decreased by continued optimization,
cf. the slope of the red line in Fig.~\ref{fig:Cs2_ass}. This is in
contrast to Fig.~\ref{fig:Cs2_symm} where $J_{leak}$ remains
essentially unchanged after about 200 iterations, suggesting that a
hard limit has been reached. Leakage from
the cooling subspace thus starts to pose a problem for
symmetrized excitation when a few hundred cooling cycles are required.
The different performance of the two optimization functionals is
not surprising since $J_T^{ass}$ is constructed to actively suppress
leakage from the initial ensemble (and the ensuing vibrational
heating) by allowing spontaneous emission only from the most favorable
instead of all accessible levels. The extent to which leakage can be
suppressed when employing $J_T^{ass}$ is nonetheless very gratifying.

\begin{figure}[tbp]
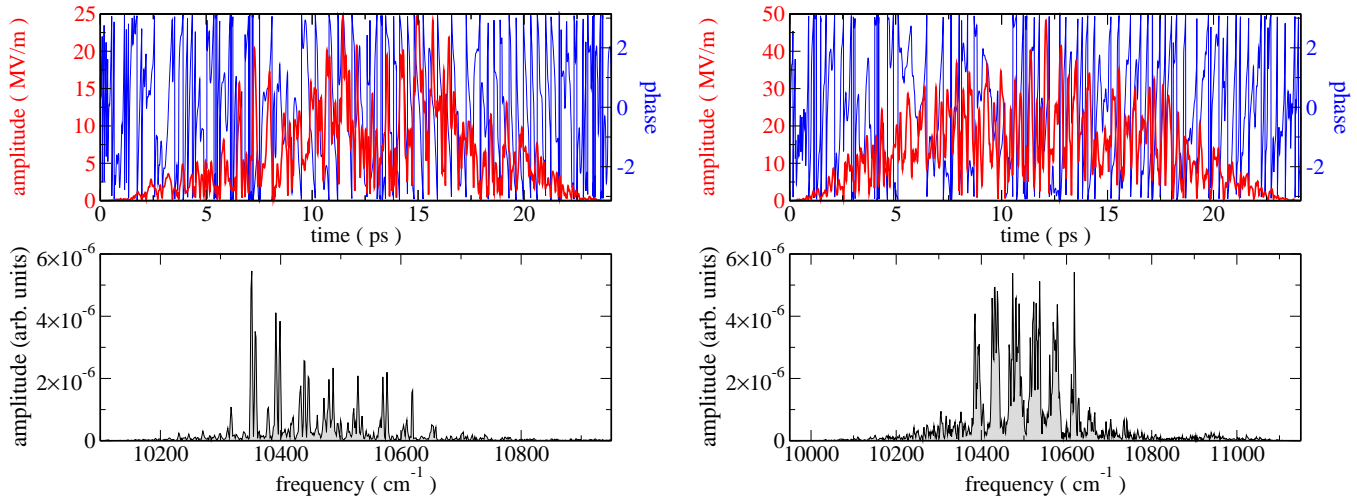

  \centering
  \includegraphics[width=0.48\linewidth]{Fig5a}
  \hspace*{0.02\linewidth}%
  \includegraphics[width=0.48\linewidth]{Fig5b}
  \caption{Optimized pulses (top) and their spectra (bottom)
    for the vibrational cooling of Cs$_2$
    molecules using symmetrized excitation (left) and assembly-line
    cooling (right).} 
  \label{fig:Cs2_pulses}
\end{figure}
The optimized pulses and their spectra for vibrational cooling of
Cs$_2$ are shown in Fig.~\ref{fig:Cs2_pulses}, comparing symmetrized
excitation (left-hand side) and assembly-line cooling (right-hand
side). The spectral width of the optimized pulses covers about
500$\,$cm$^{-1}$ corresponding to transform-limited pulses of
30$\,$fs. This is well within the standard capabilities of current
femtosecond technology. A similar conclusion can be made with respect
to the integrated pulse energies: We find $1\,\mu$J for the pulse
obtained with $J_T^{sym}$ in the left-hand side of
Fig.~\ref{fig:Cs2_pulses} and $4\,\mu$J for that obtained with
$J_T^{ass}$ in the right-hand side of Fig.~\ref{fig:Cs2_pulses}. 

We now turn to the example of LiCs molecules for which the
Franck-Condon map is not favorable to cooling. Broadband optical
pumping with unshaped pulses will thus lead to heating rather than
cooling, cf. Fig.~\ref{fig:FCF}. We demonstrate in the following that
shaping the pulses does, however, yield vibrational cooling.
Note that by employing the $B^1\Pi$-state, we have chosen the most
favorable out of all potential energy curves correlating to the lowest
excited state asymptote (Li 2s + Cs 6p). For example, the
$A^1\Sigma^+$ state is expected to be even less suited for cooling. 
While the $A^1\Sigma^+$-state potential is more 
deeply bound and could thus be somewhat better in terms of the
Franck-Condon map, it is strongly perturbed by the spin-orbit
interaction. The resulting coupling to triplet states implies a 
loss from the cooling cycle that, due to the timescale separation of
excitation and spontaneous emission, cannot be prevented by
shaping the pulse.  

\begin{figure}[tbp]
  \centering
  \includegraphics[width=0.5\linewidth]{Fig6}
  \caption{Optimizing the vibrational cooling of LiCs molecules
    using symmetrized excitation: Value of the total functional,
    Eq.~\eqref{eq:J_T_symm}, and its components vs iterations of the
    optimization algorithm ($n_{max}=5$).} 
  \label{fig:LiCs_symm}
\end{figure}
\begin{figure}[tbp]
  \centering
  \includegraphics[width=0.55\linewidth]{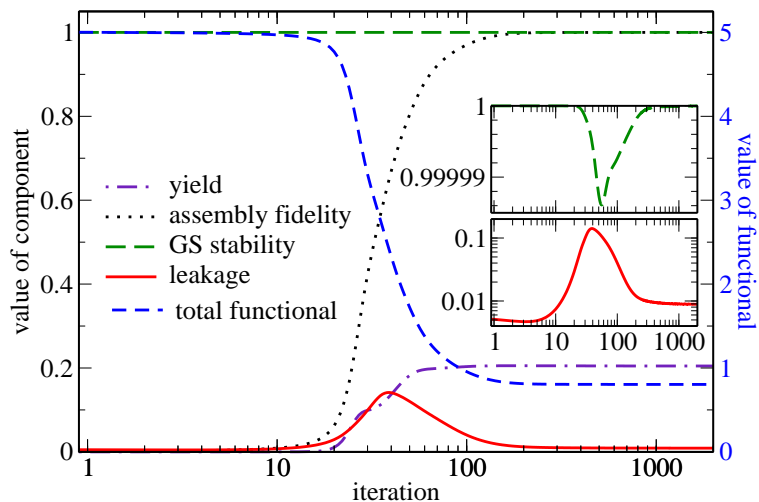}
  \caption{Optimizing the vibrational cooling of LiCs molecules
    using assembly-line cooling: Value of the total functional,
    Eq.~\eqref{eq:J_T_ass}, and its components vs iterations of the
    optimization algorithm ($n_{max}=5$).} 
  \label{fig:LiCs_ass}
\end{figure}
Since the $B^1\Pi$-state of LiCs is comparatively
shallow~\cite{GrocholaJCP09}, leakage out of the initial ensemble and
dissociation of the molecules is a more severe problem than for
Cs$_2$. We therefore first discuss $n_{max}=5$ and show later that
assembly-line cooling allows also for larger $n_{max}$. 
The behavior of the optimization functionals and their single
contributions is displayed in Fig.~\ref{fig:LiCs_symm} for $J_T^{sym}$
and in Fig.~\ref{fig:LiCs_ass} for $J_T^{ass}$. The overall behavior
of the functionals and their components is very similar to that
observed for Cs$_2$ in Figs.~\ref{fig:Cs2_symm}
and~\ref{fig:Cs2_ass}. In particular, both algorithms converge
monotonically (dashed blue lines in Figs.~\ref{fig:LiCs_symm}
and~\ref{fig:LiCs_ass}), the dark-state condition can be very well
fulfilled (green long dashed lines), and the excitation is efficient
(purple dot-dashed and black dotted lines). The behavior with respect
to leakage changes, however, dramatically when going from Cs$2$ to
LiCs (red lines in Figs.~\ref{fig:LiCs_symm} and~\ref{fig:LiCs_ass}):
$J_{leak}$ takes 
final values of 0.16 for symmetrized excitation and 0.009 for
assembly-line cooling. This reflects the Franck-Condon map being so
much more favorable to heating rather than cooling,
cf. Fig.~\ref{fig:FCF} (right), that even with shaped pulses it is
difficult to ensure cooling. In particular, the result for symmetrized
excitation is insufficient since $J_{leak}=0.16$ implies that losses from
the cooling cycle will occur already after few excitation/deexcitation
steps. For $n_{max}=5$, $J_{leak}$ reaches a plateau for symmetrized
excitation and assembly-line cooling alike. This is easily rationalized by
inspection of the Franck-Condon map in Fig.~\ref{fig:FCF} (right). In
particular, the excited state levels which are reached from $v''=5$,
such as $v'=2$, 
show a large leakage toward higher ground state vibrational levels. 
We have therefore also investigated $n_{max}=10$ for assembly-line
cooling. Most of the levels into which e.g. $v'=2$ decays and which
represent leakage for $n_{max}=5$ are then part of the
ensemble. Indeed, we find $J_{leak}=0.002$ after 1000 iterations for
$n_{max}=10$ (data not shown). Moreover, $J_{leak}$ continues to
decrease after 1000 iterations, albeit not as steeply as in
Fig.~\ref{fig:Cs2_ass} for Cs$_2$, allowing to push the value of
$J_{leak}$ below $10^{-3}$. 

\begin{figure}[tbp]
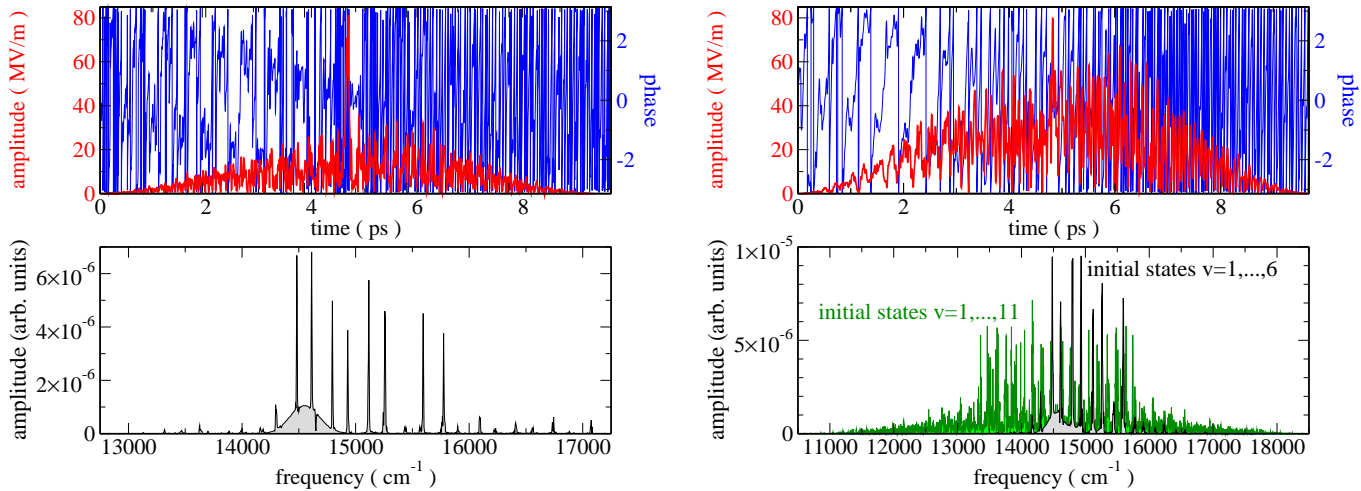

  \centering
  \includegraphics[width=0.48\linewidth]{Fig8a}
  \hspace*{0.02\linewidth}
  \includegraphics[width=0.48\linewidth]{Fig8b}
  \caption{Optimized pulses (top) and their spectra (bottom) for the
    vibrational cooling of LiCs 
    molecules using symmetrized excitation (left, $n_{max}=5$) and
    assembly-line cooling (right, $n_{max}=5$ in the top panel,
    $n_{max}=5$ and 10 in the bottom panel).} 
  \label{fig:LiCs_pulses}
\end{figure}
Figure~\ref{fig:LiCs_pulses} shows
the optimized pulses (top) and their spectra (bottom) for LiCs with 
$n_{max}=5$ and symmetrized excitation (left) and assembly-line
cooling (right). The bottom left panel
of Fig.~\ref{fig:LiCs_pulses} displays furthermore the spectrum of the
optimized assembly-line pulse obtained for $n_{max}=10$. The spectral
width obtained for $n_{max}=5$ covers less than 3000$\,$cm$^{-1}$,
corresponding to the bandwidth of a transform-limited pulse of
a few femtoseconds. The integrated pulse energy amounts to $3.4\,\mu$J. For
$n_{max}=10$, significantly more transitions need to be driven,
cf. Fig.~\ref{fig:FCF}. It is thus not surprising that both the
spectral width of the optimized pulse and its integrated energy are
larger than for $n_{max}=5$. The latter amounts to $16\,\mu$J. 
Such a pulse is more difficult to realize experimentally than those
found for Cs$_2$. The spectral width could be reduced by
employing spectral constraints~\cite{ReichKochJMO,JosePRA13}. The main 
point of our current investigation is, however, to demonstrate that
optimized pulses lead to vibrational cooling even for molecules with
unfavorable Franck-Condon map. This is evident from
Fig.~\ref{fig:LiCs_ass} and further substantiated by simulating the 
cooling process using the optimized pulses.

\begin{figure}[tbp]
  \centering
  \includegraphics[width=0.48\linewidth]{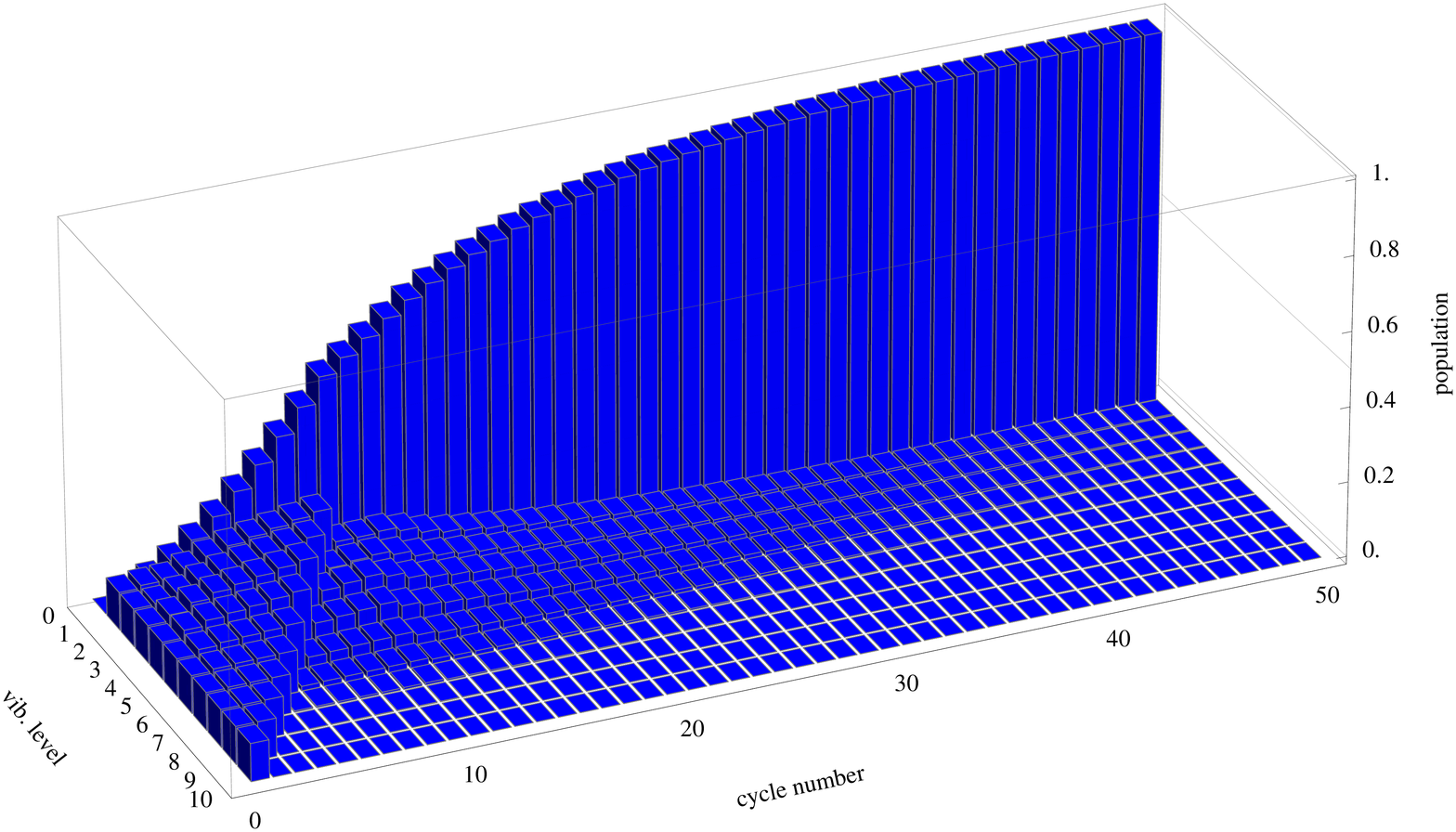}%
  \hspace*{0.01\linewidth}%
  \includegraphics[width=0.48\linewidth]{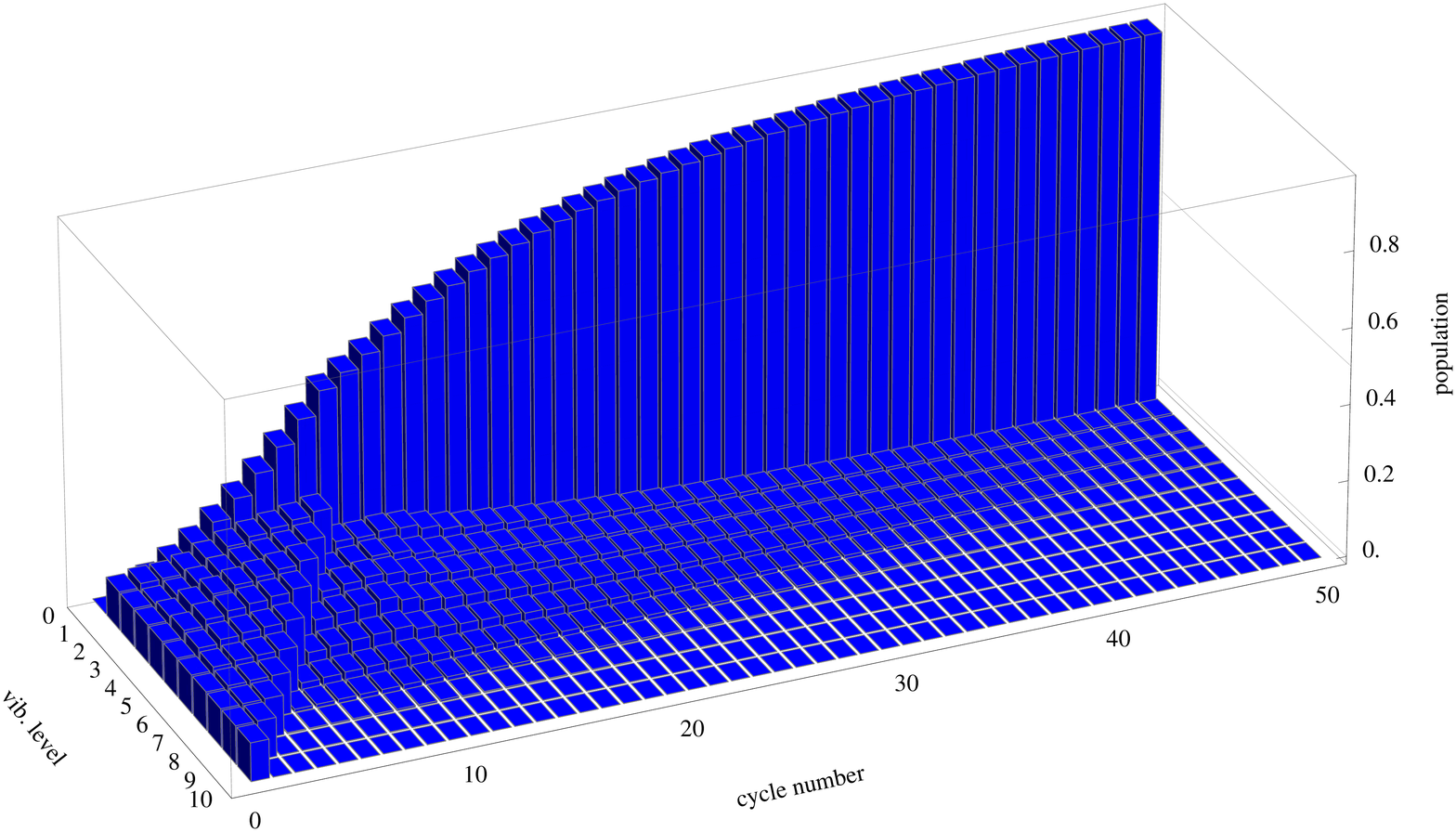}
  \caption{Demonstration of assembly-line cooling for Cs$_2$ (left)
    and LiCs (right) molecules: Population of ground state vibrational
    levels vs number of ecxitation/spontaneous emission cycles. 
    The initial distribution is assumed to
    be an equipartition in the ground state vibrational levels
    $v''=1,\ldots, v''=10$.}
  \label{fig:cooling_bars}
\end{figure}
\begin{table}[tb]
  \centering
  \begin{tabular}{|c|c|c|c|c|}
    \hline
    & cooling & no. of cycles for 90\% & max. target state yield & no. of cycles
    for max. yield \\\hline
    Cs$_2$ ($n_{max}=10$) & $J_T^{sym}$ & 23 & 0.992 & 125 \\
    Cs$_2$ ($n_{max}=10$) &  $J_T^{ass}$ & 26 & 0.9993 & 100 \\\hline
    LiCs ($n_{max}=5$) &  $J_T^{sym}$ & not achieved & 0.80 & 97 \\
    LiCs ($n_{max}=5$) &  $J_T^{ass}$ & 26 & 0.96 & 137 \\
    LiCs ($n_{max}=10$) &  $J_T^{ass}$ & 30 & 0.99 & 84 \\\hline
  \end{tabular}
  \caption{Accumulation of molecules in the target $v''=0$ level.}
  \label{tab:accumul}
\end{table}
To this end, we assume the initial incoherent ensemble to be given by 
equal population in levels
$v''=1,\ldots,10$ of the electronic ground state for both Cs$_2$ and
LiCs. We calculate the wavepacket dynamics under the optimized pulse,
and determine the ensemble that represents the initial state for the
next pulse, identical to the previous one,
by  redistributing the population according to the Einstein
coefficients, Eq.~\eqref{eq:Einstein}. The depletion of the excited
vibrational levels and accumulation of population in $v''=0$ is
imposingly demonstrated in Fig.~\ref{fig:cooling_bars} and
Table~\ref{tab:accumul}. 
A ground state population of 90\% is obtained after just a few tens of 
excitation/spontaneous emission cycles for both Cs$_2$ and LiCs.
This is in contrast to spectrally cut
pulses without any further shaping which require several thousand cycles 
for Cs$_2$ and would fail altogether for LiCs. Moreover, a high
degree of purity, $\mathcal P > 0.98$, is obtained for our optimized
pulses with only of the
order of 100 excitation/spontaneous emission cycles for both
molecules. 

\section{Summary and conclusions}
\label{sec:concl}

We have adapted optimal control theory for cooling internal degrees of
freedom to account for the timescale separation between coherent
excitation and spontaneous emission. Our approach is based on a basis
set expansion of the initial density matrix into vibrational
eigenstates. This has allowed us to carry 
optimization of vibrational cooling from toy
models~\cite{AllonJCP93,AllonJCP97,AllonCP01} to 
 a first principles description of
alkali dimer molecules that are currently studied in cooling
experiments~\cite{PilletSci08,ViteauFaraday09,SofikitisNJP09,SofikitisMolPhys10,LignierPCCP11,HorchaniPRA12,WakimOE12}.
Compared to the earlier theoretical predictions where a single long
pulse implemented the complete cooling
process~\cite{AllonJCP93,AllonJCP97,AllonCP01}, our approach allows
for finding femtosecond pulses that can be repeatedly applied, just as
is done in the experiments. Shaping the pulses using optimal control
allows to significantly reduce the number of excitation/spontaneous
emission cycles and reach a high purity of the ground state
molecules. More importantly, it also enables vibrational cooling 
for molecules where 
the Franck-Condon map favors heating rather than cooling. 

The derivation of our optimization functionals was based on two
different intuitions. First, simultaneous, symmetric excitation of all
ground state levels in the thermal ensemble to the excited state was
expected to yield most efficient cooling. It turned out, however, that
this approach has only a limited capability of suppressing leakage out
of the initial ensemble to higher lying levels. In particular for
molecules with unfavorable Franck-Condon map, this algorithm cannot
avoid vibrational heating and, in extreme cases, dissociation. We have
therefore devised an optimization functional corresponding to 
'assembly-line' cooling where only one ground state level is
transferred to the excited state while the population of all other
vibrationally excited ground state levels is reshuffled via Raman
transitions. This approach yields pulses that enforce vibrational
cooling even for molecules with transition matrix elements favoring
heating rather than cooling. The spectral widths and integrated
energies of our optimized pulses are well within the capabilities of
current femtosecond technology. We have demonstrated successful
implementation of cooling by calculating the population redistribution
over a number of excitation/spontaneous emission steps, proving
accumulation of ground state molecules. 

Our study demonstrates the power of optimal control theory for
reaching a control target that might not be accessible by simple,
analytical pulse shapes. However, it also illustrates that optimal
control theory is not a black-box tool but requires physical insight, in
particular when constructing the optimization functional. This is
crucial when one wants to address 
fundamental limits for control. In our case, this corresponds to the
question of the minimum requirement on the molecular structure that is
necessary to allow for cooling. The answer to this question determines
the controllability of the problem, irrespective of the actual
experimental resources such as pulse bandwidth or power. We find 
that all that is required is a single excited state level with
moderate spontaneous decay 
probability to the target state and a limited number of significant
transition matrix elements for the 
other ground state vibrational levels.

Laser cooling makes use of the simplest quantum reservoir, the vacuum
of electric field modes, and has led to the concept of quantum reservoir
engineering~\cite{PoyatosPRL96}. Analogously, our
optimization approach for laser cooling can be generalized to quantum
reservoir engineering. Since the creation of coherences
cannot be neglected in the general case, this requires a basis set
expansion in Liouville space rather than Hilbert space. Such a
generalization of our optimal control approach to quantum reservoir
engineering is currently in progress.

\begin{acknowledgments}
  We would like to thank Nadia Bouloufa and Olivier Dulieu for providing the
  Cs$_2$ potential energy curves. 
  The authors enjoyed hospitality of the Kavli Institute of Theoretical
  Physics at UC Santa Barbara. Financial support from the Deutsche
  Forschungsgemeinschaft (grant No. KO 2301/2) 
  and in part by the National Science Foundation (grant No. NSF
  PHY11-25915) is gratefully acknowledged.
\end{acknowledgments}


\end{document}